\documentclass [12pt,a4paper] {article}
\title {A BOHR'S SEMICLASSICAL MODEL OF THE BLACK HOLE THERMODYNAMICS }

\author{V. Pankovi\'c$^\dag$, M. Predojevi\'c$^\dag$ and P. Gruji\'c$^\S$ \\
$^\dag$Department of Physics, Faculty of Sciences, 21000 Novi Sad,\\
 Trg Dositeja Obradovi\'ca 4, Serbia,\\
$^\S$ Institute of Physics, P.O. Box 57, 11000 Belgrade, Serbia}

\date{}

\begin {document}
\maketitle

\vspace {1.5cm}
\vspace {1.5cm}

\begin {abstract}
We propose a simple procedure for evaluating the main
thermodynamical attributes of a Schwarzschild's black hole:
Bekenstein-Hawking entropy, Hawking's temperature and Bekenstein's
quantization of the surface area. We make use of the condition that
the circumference of a great circle on the black hole horizon
contains finite number of the corresponding reduced Compton's
wavelength. It is essentially analogous to Bohr's quantization
postulate in Bohr's atomic model interpreted by de Broglie's
relation. We present black hole radiation in the form conceptually
analogous to Bohr's postulate on the photon emission by discrete
quantum jump of the electron within the Old quantum theory. It
enables us, in accordance with Heisenberg's uncertainty relation and
Bohr's correspondence principle, to make a rough estimate of the
time interval for black hole evaporation, which turns out very close
to time interval predicted by the standard Hawking's theory. Our
calculations confirm Bekenstein's semiclassical result for the
energy quantization, in variance with Frasca's (2005) calculations.
Finally we speculate about the possible source-energy distribution
within the black hole horizon.

\end {abstract}
\vspace {1.5cm} Key words: black hole – entropy – quantization -
Hawking temperature \vspace {1.5cm}

\section {INTRODUCTION}
Quantum theory, both the Old one and Quantum Mechanics, was designed
to deal with microscopic phenomena, irrespective of the kind of the
interaction involved. In practice, Bohr's, Heisenberg's and
Schr$\ddot o$dinger's theories deal almost exclusively with
Coulombic interaction, as dominant at the atomic level. Strong and
weak forces are restricted to nuclear and subnuclear levels and
require a specific approach, outside the ordinary no relativistic
quantum theory, partly because these interaction are difficult to
describe by the potential functions. The fourth fundamental force,
gravitations, has been left out, for a number of reasons. First, it
appears so weak in comparison with the three ones mentioned, that at
the microscopic level can be easily ignored. Second, it is the
theory of gravitation, Newtonian, Einsteinian or else which is
considered relevant to study gravitating bodies and celestial
phenomena in general.

\subsection {Newtonian and Coulombian systems}

    Though reigning at very different scales of the physical world
    Newtonian and Coulombian forces have a common formal structure
    which makes them attribute to the corresponding physical systems
    many common characteristic (see, e.g. Gruji\'c 1993). These
    common features have been revealed in a particularly  remarkable
    way when the Quantum Mechanics was formulated and a parallel
    with some General Relativity phenomena established.

    The first modern hydrogen atom model was contrived by Thomson as
    the negatively charged electron immersed in a spherical positively charged
    fluid. This model was to be radically changed with the later
    Rutherford-Bohr model, but both had one remarkable feature in
    common: the path an electron traced while moving inside the fluid, or
    around the nucleus, was the same geometrical figure, an ellipse, despite the fact
    that electron experiences radically different forces. In the Thomson's model, the
    potential function is of the form

    \begin{equation}
            V(r) = k r^2
    \end{equation}
    \noindent
    that is as for the harmonic oscillator, whereas for the Coulombic  interaction one has

    \begin{equation}
            V(r) = \beta/r
    \end{equation}

    The difference was the positions of their foci. In the first case they
    were placed symmetrically with respect to the centre of the sphere, whereas
    in the motion around point-like nucleus the latter was positions at one of
    two foci. But the most remarkable similarity was revealed when comparing
    the semiclassical and quantum mechanical solutions of the corresponding energy
    spectra. It turns out hat in both cases semiclassical and quantum mechanical
    results coincide, for all principal quantum numbers (so-called {\it
    correspondence identities}) (see, e.g. Norcliffe 1975). Hence, two most important
    interactions, harmonic oscillators and Coulombic ones allow for the semiclassical
    and quantum mechanical treatments indiscriminately. At the same time, it is for these
    interactions that any single-particle trajectory is closed, irrespective of
    the initial conditions.  Not accidentally, these (textbook) interactions appear
    the only that allow for exact analytical solutions, both classical and quantum mechanical.

    The energy spectra, evaluated in either of the theories, appear
    distinct, however. For the harmonic oscillator from Eq. (1) it
    reads

    \begin{equation}
    E_n = \hbar \omega(n+3/2), n = 0,1,2,...
    \end{equation}
    \noindent
    whereas the Coulombic case provides

    \begin{equation}
    E_n = \beta^2/2n^2, n = 1,2,...
    \end{equation}
    \noindent
    The first formula provides an equidistant distribution, whereas the
    Bohr's formula is the typical case for a series of discrete levels
    accumulating towards zero. It is this distinction which makes
    the investigation of the black hole spectrum of a particular
    interest, as we shall see below.

\subsection{Black hole}

It was only with the appearance of the concept of gravitational
collapse and the model of black hole that the gravitational force
becomes dominant and even exclusively present (se, e.g. Bekenstein
1994).  In view of the formal similarity of the asymptotic behaviour
of the Newtonian and Coulombic forces one may expect that the
properties of atomic systems with charged constituents and
black-hole like gravitational objects should share a number of
common features. As noted by Bekenstein (1998) black hole is a
hydrogen atom in the field of the strong gravity regime. In
particular quantum effects may be present on the black hole surface
and one may expect that some quantization rules are applicable.

         One of the essential ingredients of the statistical
         mechanics has been the observation that the number of degrees of
         freedom of a quantum system should be proportional to the
         surface of the system, rather than to  the volume.  In fact,
         it was this assumption which led to the Bekenstein's linking of the
         black hole entropy and the area of its horizon.

        Semiclassical quantization of black hole (BH) has been
        attempted by various authors. In a recent work Frasca (2005)
        calculated the semiclassical energy spectrum of the Schwarzschild black
        hole making use of the Hamilton-Jacobi formalism. For the
        stable circular orbits he derived the formula

\begin{equation}
E_n \approx M - {2G^2M^5\over n^2\hbar^2}
\end{equation}
\noindent which is, up to a additive constant M, Bohr's formula for
the Coulombic interaction. In addition, the partition function
turned out to coincide with that derived by the loop quantum gravity
formalism (see, e.g. Nicolai \emph{et al} 2005).

\subsection {Black hole characteristics}

Thermodynamical characteristics of a black hole is one of most
important subject of the contemporary physics (see, e.g. very recent
paper by Samuel and Chowdhury 2007). In a sense it plays the role of
the black body studies around the turn of $19^{th}$ century, linking
the thermodynamics and statistics, more precisely the gravitation
and information theories. First, Bekenstein (1973) suggested that a
black hole contains the entropy $S_{BH}$ proportional to the horizon
surface area, $A$. For the Schwarzschild's black hole one has:
\begin {equation}
     S_{BH} = \frac {k_{B}c^{3}}{4\hbar}A
\end {equation}
\noindent where $ k_{B}$  is Boltzmann's constant, $c$ – speed of
light and $\hbar$ – reduced Planck constant. Also, Bekenstein
suggested that the horizon surface area is quantized, and can  be
changed only discretely
\begin {equation}
     \Delta A = n 8\frac{G\hbar}{c^{3}} \equiv n 8L^{2}_{P} ,      n = 1, 2, …      .
\end {equation}
\noindent where $L_{P}=(\frac{G\hbar}{c^{3}})^{\frac {1}{2}}$ is the
Planck's length. Bekenstein's analysis is, on the one hand, based on
the characteristics of corresponding, complex quantum measurement
procedures, i.e. Heisenberg's uncertainty relations and Ehrenfest's
adiabatic theorem. On the other hand it relies on general
relativistic and quantum field theoretical requirement on the
stability of the capture of a quantum system within black hole.
According to this requirement, roughly speaking, Compton's
wavelength of a given quantum system must be smaller than double
Schwarzschild's radius. (Otherwise a quantum system can escape from
black hole by means of quantum tunneling.)

After Bekenstein, Hawking (1975, 1976) showed that black hole can be
considered as a black body which radiates at the temperature
\begin {equation}
      T_{H} = \frac {\hbar c^{3}}{8\pi k_{B}GM}
\end {equation}
where $M$ is the black hole mass. This Hawking's temperature,
according to usual rules of the thermodynamics, appears compatible
to Bekenstein-Hawking entropy (6). Roughly speaking, Hawking's
analysis is physically based on the non-invariance of the quantum
field dynamics according to general transformations of coordinates,
which implies that a  wave can be considered as a complex mixture of
the plane waves (this mixture can be effectively treated as the
spectrum of the black body). Simplified, black hole can
gravitationally interact with fluctuated quantum vacuum near
horizon. Then black hole can absorb one member of
particle-antiparticle virtual pair, while other member of the pair
can be effectively considered as the radiation. Mathematically,
Hawking's analysis is based on a complex formalism of the quantum
fields in the curved space (in a quasi-classical approximation).
Later it has been proved, by Hawking (1979) and others, that
Hawking's results can be reproduced even by more complex formalism,
i.e. quantum field dynamics without quasi-classical approximations
(see, for example, review articles (Wald 1997, 1999, Page 2004) and
references therein).

Hawking, also, predicted time of the black hole evaporation.
    Namely, Stefan-Boltzmann law applied at the black hole surface area,
     according to Hawking's temperature (9) and relativistic equivalence
     relation $E = M c^2$, has the form

\begin {equation}
-dE/dt = - c^2 dM/dt = \sigma_{SB} T_H^4 A = \frac{hc^6}{15360 \pi
G^2 M^2},
\end {equation}
\noindent where $\sigma_{SB}$ is the Stefan-Boltzmann constant. It
yields, after simple integration, the following expression for black
hole evaporation
\begin {equation}
\tau_{ev} = 5120 \frac{\pi G^{2}}{\hbar c^{4}}M^{3}_{0}
\end {equation}
where $M_{0}$ denotes the black hole initial mass.

Further, detailed analysis of the quantum and thermodynamical
characteristics of a black hole needs a very complex (in this moment
incomplete) theoretical formalism including application of different
string theories (Strominger and Vafa 1996, Proline 2006).
Nevertheless, there are many attempts of the analysis of quantum and
thermodynamical characteristics of a black hole by relatively simple
(approximate) theoretical concepts. For example, in (Ram 2000, Ram
{\it et al} 2005) it is shown that a black hole can be consistently
considered as a Bose-Einstein condensate, while in (Nagatani 2007) a
conceptual analogy between so-called minimum black hole and Bohr's
model of the hydrogen atom is considered. Even in these cases
mathematical formalism is based on different differential (e.g.
Schr$\ddot o$dinger's) equations solved within some (e.g. mean
field) approximations.

\section {THEORY}

In this work we shall determine, in a simple way, three most
important, thermodynamical characteristics of a Schwarzschild's
black hole: Bekenstein-Hawking's entropy, Hawking's temperature and
Bekenstein's quantization of the surface area. We shall use an
original, simple and intuitively (quasi-classical) transparent
condition. We demand that circumference of a great circle at black
hole horizon contains integer (statistically averaged) number of
corresponding reduced Compton's wavelength. It is essentially
analogous to Bohr's quantization postulate in his Old quantum
theory, interpreted by de Broglie's ontology, according to which
circumference of an electron circular orbit comprises an integer
number of corresponding de Broglie's wavelengths. Finally, we
express the black hole radiation in the form conceptually analogous
to Bohr's postulate on the photon emission by discrete quantum jump
of the electron in his atomic model. It, in accordance with
Heisenberg's energy-time uncertainty relation and a correspondence
principle conceptually analogous to Bohr's one, admits a rough
estimate of the time interval for black hole evaporation. This time
interval is very close to the time interval of the black hole
evaporation obtained via Hawking's radiation.

Thus, in this work we shall make a most simplified but non-trivial
description of the quantum and thermodynamical characteristics of a
Schwarzschild's black hole, which we simply call Bohr's black hole.

\subsection{Bohr's black hole}

Making use of de Broglie's relation
\begin {equation}
        \lambda = \frac {h}{mv}
\end {equation}
and Bohr's quantization postulate
\begin {equation}
          mv_{n}r_{n} = n\frac {h}{2\pi},           n = 1, 2,...
\end {equation}
it follows
\begin {equation}
          2\pi r_{n} = n \lambda_{n}  , n = 1, 2, ...
\end {equation}
where $\lambda_{n}$  represents the $n$-th electron de Broglie's
wavelength, $m$ – electron mass, $v_{n}$ – electron $n$-th speed,
$r_{n}$ - radius of the electron $n$-th circular orbit and $h$ –
Planck constant. Expression (13) simply means that circumference of
$n$-th electron circular orbit contains exactly $n$ corresponding
$n$-th de Broglie's wavelengths, for $n$ = 1, 2, ...

We shall now apply similar analysis of a Schwarzschild's black hole
with mass $M$ and Schwarzschild's radius
\begin {equation}
             R_{S} = {2GM\over c^2}                   .
\end {equation}
We introduce the following expression analogous to (12)
\begin {equation}
            m_{n}c R_{S} = n\frac {\hbar}{2\pi},  n = 1, 2, ...
\end {equation}
what implies
\begin {equation}
             2\pi R_{S} = n \frac {\hbar}{m_{n}c},  n = 1, 2, ...
             \end {equation}
analogous to (9). Here $2\pi R_{S}$ represents the circumference of
the black hole while
\begin {equation}
           \lambda_{r n}= \frac {\hbar}{m_{n}c}, n = 1, 2, ...
           \end {equation}
\noindent
 is $n$-th reduced Compton's wavelength of a quantum system
captured at the black hole horizon surface for expression (16)
simply means that {\it circumference of the black hole horizon holds
exactly $n$ reduced Compton's wave lengths of a quantum system
captured at the black hole horizon surface}. Obviously, it is
essentially analogous to above mentioned Bohr's quantization
postulate interpreted via de Broglie's relation. However, there is a
principal difference with respect to Bohr's atomic model. Namely, in
Bohr's atomic model different quantum numbers $n = 1, 2, ...$ ,
correspond to different circular orbits (with circumferences
proportional to $n^{2}$). Here any quantum number $n = 1, 2, ...$
corresponds to the same circular orbit (with circumference $2\pi
R_{S}$).

According to (11) it follows
\begin {equation}
           m_{n} = n\frac {\hbar}{2\pi c
            R_{S}} = n \frac {\hbar c}{4\pi GM}\equiv n m_{1} , n = 1, 2, ...
\end {equation}
where
\begin {equation}
          m_1 = {\hbar c\over 4\pi GM} = {M_P^2\over 4\pi M}
\end {equation}
\noindent and where $M_{P}= (\hbar c/G)^{1/2}$ is the Planck mass.
Obviously, $m_{1}$ depends of $M$ so that $m_{1}$ decreases when $M$
increases and vice versa. For a macroscopic black hole, i.e. for $M
\gg M_{P}$ it follows $m_{1} \ll M_{P}$.

Suppose now that black hole mass equals
\begin {equation}
           M = \sigma m_{1} = {\sigma \hbar\over 4\pi cR_S} = {\sigma \hbar c\over 4\pi GM}
\end {equation}
\noindent where $\sigma$ denotes some integer (or approximately
integer) number. According to (14,18) it follows
\begin {equation}
         \sigma = {M\over m_1} = {4\pi GM^{2}\over \hbar c}           .
\end {equation}
It means that the number $\sigma$, for fixed black hole mass $M$, is
finite.

After multiplication of (17) by Boltzmann constant $k_{B}$ we have

\begin {equation}
           k_{B}\sigma  = {4\pi k_{B}GM^{2}\over \hbar c}
\end {equation}
Obviously, right-hand side of (18) represents Bekenstein-Hawking's
entropy of the Schwarzschild's black hole (6). It is therefore
reasonable to assume
\begin {equation}
S_{BH} = k_{B}\sigma = \frac {4\pi k_{B}GM^{2}}{\hbar c}
\end{equation}

This assumption implies that $\sigma$ must have a statistically
appropriate form that will be considered later on.

Differentiation of (23) yields

\begin {equation}
dS_{BH} = k_{B}d\sigma = 8\pi k_{B}GM/(\hbar c^3) dE
\end {equation}

\noindent
where
\begin {equation}
          E=Mc^{2}
\end {equation}
\noindent

 is the black hole energy. Expression (24), according to
first thermodynamical law, implies that term

\begin {equation}
         T = {\hbar c^{3}}/{8\pi k_{B}GM} = m_{1}c^{2}/(2k_{B})
\end {equation}
\noindent represents the black hole temperature. Evidently, this
temperature is identical to Hawking's black hole temperature (8).
According to (23), (24) it follows
\begin {equation}
        dA = \frac {32\pi G^{2}}{c^{3}} M dM
\end {equation}
\noindent
or, in a corresponding finite difference form
\begin {equation}
        \Delta A = \frac {32\pi G^{2}}{c^{3}} M \Delta M ,     for   \Delta M \ll M .
\end {equation}
Further, we assume
\begin {equation}
            \Delta M = m_{n} –m_{k} = (n-k) m_{1} = \frac {\hbar c}{4\pi GM},   n,k < n = 1, 2, …
\end {equation}
\noindent which, after substituting in (28), yields
\begin {equation}
         \Delta A_{n k} = (n-k) 8\frac {G\hbar}{c^{3}} = (n-k) 8L^{2}_{P} , (n-k) = 1, 2, …            .
\end {equation}
Obviously, expression (30) represents Bekenstein's quantization of
the black hole horizon surface area (7).

In this way we have reproduced, i.e. determined in an independent
way, three most important characteristics of Schwarzschild's black
hole thermodynamics: Bekenstein-Hawking entropy, Hawking's
temperature and Bekenstein's quantization of the surface area.

   We now evaluate the necessary statistical form of $\sigma$. We suppose that black
hole can be considered as a canonical statistical ensemble of
Bose-Einstein quantum systems. Then the statistical sum, Z,
according to (18), equals
\begin {equation}
          Z = \sum_{n=0}\exp[- \frac {E_{n}}{k_{B}T_{H}}]
\end {equation}
\noindent
where
\begin {equation}
          E_{n} = m_{n}c^{2} = n \frac {\hbar c^{3}}{4\pi GM}
          = n \frac {M_{P}}{M}\frac {E_{P}}{4\pi} = n E_{1} , n = 0, 1, 2, ....
\end {equation}
\noindent and where $E_{P}= M_{P}c^{2}$ is Planck energy. (It is
supposed, implicitly, that $n$ can be zero. Or, precisely, it can be
shown  by a more detailed analysis, $n$ can be changed by
$(l(l+1))^{\frac {1}{2}}$ for $l =0, 1, ...$)

Hence, our calculations provide harmonic-oscillator-like spectrum,
supporting Bekenstein's result, and in variance with Frasca's (2005)
calculations. The difference between the latter two approaches
concerns not only the mere spectrum of the black hole energy, but
may shed the light onto the possible spatial energy-distribution
within black hole. As described above, equidistant energy level
distribution signals a uniform source-matter distribution, as the
case of Thomson's atomic models shows.  If the
harmonic-oscillator-like spectrum proves correct, this would imply
the uniformity of the gravitational field within the horizon.

According to (8), (33), it follows
\begin {equation}
                \frac {E_{1}}{k_{B}T_{H}} = 2
\end {equation}
\noindent
 which introduced in (32) yields
\begin {equation}
  Z = \sum_{n=0}\exp[- 2n] = \exp[2]/(\exp[2]-1)
\end {equation}
\noindent
Then
\begin {equation}
             w_{n} = \exp[- \frac {E_{n}}{k_{B}T_{H}}]/Z =
             {(\exp[2]-1)}\frac {\exp[-2n]}{\exp[2]}
\end {equation}
\noindent represents probability of quantum (energy eigen) state
$n$.

Further, it follows
\begin {equation}
        M = -c^{-2}\frac {\partial (ln[Z])}{\partial (1/(k_{B}T_{H}))}= \sum_{n=0} w_{n}m_{n} = \sum_{n=0} w_{n} n m_{1}= m_{1}\sum_{n=0} w_{n} n
\end {equation}
\noindent
 which implies
\begin {equation}
            \sigma = \sum_{n=0}w_{n} n  = <N>            .
\end {equation}
Obviously $\sigma$ can be considered as the statistical average
value $<N>$ of the number of the quantum (energy eigen) states. On
the other hand $\sigma$ considered as statistically determined
entropy (in $k_{B}$ units) must have form
\begin {equation}
                \sigma = -\sum_{n=0} w_{n}ln[w_{n}]
\end {equation}
Consistency of the analysis needs that (38) and (39) be equivalent
which implies that condition
\begin {equation}
             n = ln[w_{n}] , n = 0, 1, 2, ...
\end {equation}
\noindent
 must be satisfied. However, according to (35), it follows
\begin {equation}
              ln[w_{n}] = 2n – ln[\frac {\exp[2]-1}{\exp[2]}] \approx 2n  - 0.145 , n = 0, 1, 2, ...
\end {equation}
\noindent

what reveals that condition (39) is not satisfied. Nevertheless we
note that left- and right-hand sides of (39) have the same order of
magnitude, precisely that for large $n$ right hand of (38) is twice
greater than left-hand side of (41),what appears an interesting
result.

We assume now that black hole represents a great statistical
ensemble of Bose-Einstein systems with statistical sum
\begin {equation}
          Z = \sum_{n=0}\exp[- \frac {E_{n}-\mu n}{k_{B}T_{H}}]
\end {equation}
\noindent
 where $\mu$ represents the chemical potential while $n$
in$\mu n$ can be considered as statistical average value of
Bose-Einstein systems in quantum (energy eigen) state $n$.

Suppose, further,
\begin {equation}
           \mu = \frac {E_{1}}{2}
\end {equation}
\noindent
 which, according to (32), implies
\begin {equation}
            \mu n = \frac {E_{n}}{2} , n = 0, 1, 2, ...
\end {equation}
\noindent and, according to (32)
\begin {equation}
                   Z = \sum_{n=0}\exp[- \frac {E_{n}}{2k_{B}T_{H}}].
\end {equation}
In (44) $Z$ can be considered as the statistical sum of a canonical
ensemble with
\begin {equation}
w_{n}= \exp [-E_{n}/{2k_{B}T_{H}}] (\exp[1]-1)
\exp[-n]/(Z\exp[1]), n = 0, 1, 2, ...
\end {equation}
\noindent

 It implies
\begin {equation}
         \ln[w_{n}] = n – \ln[\frac {\exp[1]-1}{\exp[1]}] \approx n  - 0.45 ,  n = 0, 1, 2, ...
\end {equation}
\noindent
and
\begin {equation}
              \ln[w_{n}] \approx n, n \gg 1              .
\end {equation}
Relation (47) implies that condition (39) concerning the consistent
statistical interpretation of $\sigma$ is well satisfied for
probabilities effectively defined by (46).

\subsection{Energy spectrum and evaporation time}

  In Bohr's atomic model we have the postulate on the energy emission
  by discrete, spontaneous, quantum jump of the electron from a higher onto a lower
  circular orbit. This quantum jump represents an effective final result
  (or simplified description) of the electromagnetic self-interaction of
  the atom. Also, according to Bohr's correspondence principle, emission
  of the photon appears most probably by quantum jump of the electron from
  an initial, sufficiently high quantum state $n$ onto the neighbouring final
  quantum state $(n-1)$.

 In conceptual analogy with Bohr's atomic model, suppose
that black hole, considered as Bose-Einstein quantum system, in some
initial quantum state $n$ can spontaneously and discretely (by means
of gravitational self-interaction) pass, i.e. jump, to some final,
lower quantum state $k$, for $k<n=1, 2,...$. Suppose, also, that by
this quantum jump an effective final emission of a quantum of energy
takes place which propagates far away from the black hole. Of
course, black hole, according to its classical definition, captures
any physical system near horizon by means of the gravitational
interaction. Nevertheless, {\it according to principles of the
quantum theory, (quantum mechanics and quantum field theory alike)
gravitationally self-interacting black hole passes from an initially
non-stable quantum state} $n$ in the final, stable quantum state
$k<n$ {\it by emitting one energy quantum outside horizon}. This is,
of course, a simplified, phenomenological description of the black
hole gravitational self-interaction.

Energy of given energy quantum, according to (32), equals
\begin {equation}
E_{n}-E_{k}=  \hbar \omega_{nk} , k < n = 1, 2, ...
\end {equation}
\noindent
 where $\omega_{nk}$ is the circular frequency of given
energy quantum. Then, according to (33), it follows
\begin {equation}
E_{n}-E_{k}= E_{n-k}=(n-k) {c^3\over 4\pi GM}, k < n=1,2,...
\end {equation},

Here we assume that a correspondence principle, conceptually similar
to Bohr's, holds. Precisely, suppose that for initial, large quantum
state $n$, there is most probable quantum jump to the final state
$k=n-1$, with corresponding emission of the one energy quantum
\begin {equation}
E_{n}-E_{n-1} =  E_{1} = {\hbar c^3\over 4\pi GM} , n = 1, 2,...
\end {equation}

Of course, given quantum jump can be considered definitive, i.e.
irreversible, if and only if condition
\begin {equation}
\Delta E_{n} + \Delta E_{n-1} \ll E_{n}- E_{n-1} =  E_{1}, n =
1,2,...
\end {equation}
\noindent
is satisfied. Here $\Delta E_{n}$ and $\Delta E_{n-1}$
represent the energy natural widths of quantum states $n$ and $n-1$
and, for sufficiently large $n$ we assume
\begin {equation}
               \Delta E_{n} \approx \Delta E_{n-1}
\end {equation}
For a more accurate form of (51) a more rigorous form of the quantum
gravitation is necessary. Nevertheless, we shall simply suppose,
according to (52),
\begin {equation}
             2\Delta E_{n}\leq \frac {E_{1}}{100} , n \gg 1 .
\end {equation}
According to Heisenberg's energy-time uncertainty relation
\begin {equation}
        \tau \Delta E_{n}\approx \frac {\hbar}{2}  , n \gg 1,
\end {equation}
\noindent
 where $\tau$ represents the time of the one energy quantum
emission or life time of the Bose-Einstein system in the initial
quantum state, it follows
\begin{equation}
  \Delta E_n \approx \hbar/(2\tau) , n \gg 1                         .
\end{equation}

Then, according to (53), (55), it follows
\begin{equation}
 t = 100 \hbar/E_1 =100 \cdot 4\pi GM/c^3 ,  n \gg 1                         .
\end{equation}

Suppose now that a black hole is initially in the (statistically
averaged) quantum state $\frac {M}{m_{1}}$. Let the black hole,
according to previous discussion, emit by quantum jump, energy
quantum $E_{1}$ within time interval $\tau$. It implies that initial
black hole with mass $M$ will entirely evaporate by means of its
gravitational self-interaction after a time interval $\tau_{ev}$.
Given time interval can be roughly estimated, according to (19),
(32), by

\begin {equation}
\tau_{ev} \geq \frac {M}{m_{1}}100 \frac {\hbar}{E_{1}} =100 (16\pi)
\frac { \pi G^{2}M^{3}}{c^{4}}\approx 5027 \frac { \pi
G^{2}M^{3}}{c^{4}}.
\end {equation}

We see that the result is very close to Hawking's time for black
hole evaporation (10).

 \section {CONCLUSION}

 We have carried out a simplified but non-trivial quasi-classical analysis
 of the quantum and thermodynamical characteristics of a Schwarzschild's black
 hole. Our analysis is conceptually analogous to formalism of Bohr's atomic model and for
this reason  black hole in our description can be simply called
Bohr's black hole. We started by a condition, analogous to Bohr's
quantization postulate, via de Broglie relation. This condition
states that circumference of a great circle at black hole horizon
contains an integer (statistically averaged) number of corresponding
reduced Compton's wavelength. It implies simple determination of
three most important thermodynamical characteristics of black hole:
Bekenstein-Hawking entropy, Hawking's temperature and Bekenstein
quantization of the surface area. Finally, we presented black hole
radiation in the form conceptually analogous to Bohr's postulate on
the photon emission by discrete quantum jump of the electron in
Bohr's atomic model. It, in accordance with Heisenberg's energy-time
uncertainty relation and a correspondence rule conceptually
analogous to Bohr' s correspondence principle, admits a rough
estimate of the time interval for black hole evaporation. This time
interval is very close to time interval of the black hole
evaporation obtained via Hawking's radiation.

   Finally, we have speculated about the relevance of the energy
   spectrum for the evidence of the source-field distribution within the horizon,
   which is, otherwise, unobservable quantity.

\section {REFERENCES}

 Bekenstein, J. D.: 1973, {\it Phys. Rev.}, {\bf D7}, 2333.

 Bekenstein,  J. D.: 1994, arXiv:gr-qc/9409015 v2 12 Sep 94.

 Bekenstein,  J. D.: 1998, arXiv:gr-qc/9808028 v3 2 Nov1998..

 Pavon,  D.: 2007, {\it J. Phys. A: Math. Theor.} {\bf 40},6865-6869.

 Frasca, M.: 2005, arXiv:hep-th/0411245 v4 10 Aug 2005..

 Grujic, P. V.: 1993, {\it Bull. Astron.}, {\bf 147}, 15-29.

 Hawking, S. W.: 1975, {\it Comm. Math. Phys.}, {\bf 43}, 199

 Hawking,  S. W.: 1979, in {\it General Relativity, an Einstein Centenary Survey},
           Eds. S. W. Hawking, W. Israel (Cambridge University Press, Cambridge           )

 Nagatani, Y.: 2007, {\it Progr. Theor. Phys. Suppl.},{\bf 164}, 54;
 hep-th/0611292.

 Nicolai,H., Peeters, K. and Zamaklar, M.: 2005, {\it Class. Quantum
             Grav.}, {\bf 22}, R193-R247.

Norcliffe A.: 1975, in {\it Case Studies in Atomic Physics}, Vol. 4,
Eds E. W. McDaniel and M. R. McDowell (North-Holland, Amsterdam),
pp. 46-55.

 Page, D. N.: 2004, {\it Hawking Radiation and Black Hole Thermodynamics}, hep-th/0409024

 Proline,  B.: 2006, {\it Lectures on Black Holes,Topological Strings and
 Quantum Attractors}, hep-th/0607227.

 Samuel, J. and  Chowdhury, S. R.: 2007, {\it Class. Quantum Grav.}, {\bf 24}, F47.

 Strominger, D. N., Vafa, C.: 1996, {\it Phys. Lett.} {\bf B 339}, 99 ; hep-th/9601029.

 Ram, B.: 2000, {\it Phys. Lett.}, {\bf A 265}, 1.

 Ram, B., Ram, A, Ram, N.: 2005, {\it The Quantum Black Hole}, gr-qc/0504030.

 Wald,  R. M.: 1997, {\it Black Hole and Thermodynamics}, gr-qc/9702022

 Wald, R. M.: 1999, {\it The Thermodynamics of Black Holes}, gr-qc/9912119

\eject\null

\vskip0.8cm
\begin{center}
{\Large BOROV SEMICLASICAN MODEL TERMODINAMIKE CRNE RUPE}
\end{center}

\vskip0.5cm
\begin{center}
  {\large V. Pankovi\'c$^1$, M. Predojevi\'c$^1$

  i P. Gruji\'c$^2$} \\
  {\large

  $^1$Katedra za fiziku, Prirodno-matemati\^ cki fakultet, 21000
Novi Sad,\\ Trg Dositeja Obradovi\'ca 4, Srbija,\\
$^2$Institut za fiziku, P.F. 57, 11000 Beograd, Srbija\\}
\end{center}

Predlo\^ zen je jednostavan postupak za izra\^ cunavanje osnovnih
termodinami\^ ckih atributa \^ Svarc\^ sildove crne rupe: Beken\^
stajn-Hokingove entropije, Hokingove temperature i Beken\^ stajnove
kvantizacije povr\^ sine horizonta. Kori\^ s\'cen je uslov da obim
velikog kruga horizonta sadr\^ zi ceo broj redukovane Komptonove
talasne du\^ zine. Postupak je analogan Borovom postulatu za
kvantizaciju atoma vodonika preko de Broljeve relacije. Postupak
implicira uobi\^ cajeno zna\^ cenje entropije crne rupe, u odnosu na
povr\^ sinu kvantne varijacije velikih krugova na horizontu. Zra\^
cenje crne rupe prezentirano je u obliku analognom Borovom konceptu
emisije fotona putem diskrenih kvantnih skokova u okiru Stare
Kvantne teorije. To omogu\'cava, prema Hajzenbergovim relacijama
neodredjenosti i Borovom principu korespodencije, procenu vremenskog
intervala za isparenje crne rupe, za koje je nadjeno da je veoma
blisko intervalu prema standarardnoj Hokingovoj formuli. Najzad,
diskutovane su posledice izra\^ cunate energijske raspodele  na
procenu raspodele energije unutar crne rupe.

\end {document}